\documentclass[preprint,showpacs,preprintnumbers,amsmath,amssymb]{revtex4}

\usepackage{graphicx}
\usepackage{dcolumn}
\usepackage{bm}
\usepackage{epsfig}
\usepackage{subfigure}
\def\e{\begin{equation}}
\def\f{\end{equation}}
\def\=#1{\overline{\overline #1}}

\def\_#1{{\bf #1}}
\def\.{\cdot}

\def\o{\omega}

\def\va{\varepsilon}

\begin{document}

\title{On cavity-and-surface enhanced Raman scattering from metamaterial shells}

\author{Constantin R. Simovski
}
\affiliation{
Department of Radio Science and Engineering/SMARAD, Helsinki
University of Technology, P.O. Box 3000, FI-02015 TKK, Finland
}

\begin{abstract}
In this paper we theoretically show that the Raman scattering by a
core-shell micron or submicron particle with epsilon-near-zero
metamaterial shell and silica spherical or cylindrical core can
combine useful features of cavity-enhanced and surface-enhanced
Raman scattering. The cavity resonance together with the plasmon
resonance lead to the giant enhancement of the field inside the
metashell which is performed as a layer of silver or gold
nanoparticles and is penetrable for molecules to be detected. This
approach results in the significant increase of both effective
volume in which molecules are affected by enhanced electric field
and Raman gain averaged over this volume.
\end{abstract}

\pacs{78.45.1h, 78.30.Er, 78.30.Fs, 78.67.Bf, 78.55.Et, 73.20.Mf,
85.352p,  05.45.-a, 07.79.Fc, 42.55.Sa}

\maketitle

The role of the surface-enhanced Raman scattering (SERS)
\cite{SERS1} in the modern sensing, especially in molecular
detection, is huge. The mechanism of SERS is related with
plasmonic nanoparticles resulting (in conventional SERS schemes)
from roughening the silver interface. The nanoparticles offer the
resonant enhancement of the local field acting on molecules
(located near them in a liquid or gas host medium) compared to the
incident wave field. This enhancement corresponds to the
proportional increase of the molecule dipole moments at the
excitation frequency $\o_e$ and at Raman frequencies $\o_R$ (e.g.
\cite{SERS11}). This effect is complemented by the similar
increase of the radiation of the pair molecule plus nearest
nanoparticle at the Raman frequency (e.g.
\cite{SERS11,SERS_exp1,SERS_exp2}). The plasmon resonance
experienced by a plasmonic nanoparticle (sphere or ellipsoid) is
rather wide-band. Therefore, usually, one of Raman frequencies
radiated by molecules or both of them together with the excitation
frequency lie within the nanoparticles resonance band.
Practically, what is detected in SERS is not the radiation of
molecules  at frequency $\o_R$ but the re-radiation of resonant
nanoparticles excited by molecules at $\o_R$. For the amplitude of
the field scattered by the pair molecule plus nanoparticle
enhanced due to the presence of the nanoparticle the coefficient
$\kappa(\o_e)$ expressing the local field enhancement is
multiplied by the coefficient $\chi(\o_R)$ which equals to the
ratio of dipole moments of a resonant nanoparticle to that of a
molecule. In the classical theory
\cite{SERS11,SERS_exp1,SERS_exp2} it is shown that
$\kappa(\o)=\chi(\o)$. When the Raman shift is small i.e.
$\o_R\approx \o_e\approx \o_{\rm av}\equiv \o_e/2+\o_R/2$ this
result corresponds to the SERS amplitude electromagnetic gain
$G_{\rm max}=\kappa_{\rm max}(\o_e) \chi_{\rm max})(\o_R)\approx
\kappa_{\rm max}^2(\o_{\rm av})$ of the order $10^{(1.5-2.5)}$ or
to the SERS intensity electromagnetic $GI_{\rm max}=\kappa_{\rm
max}^4(\o_{\rm av}) \sim 10^{(3-5)}$ \cite{SERS11,SERS_exp_add}.
Here the subscript corresponds to the optimal location of a
molecule with respect to the nanoparticle and the incident wave
vector because only the radial (with respect to the nanoparticle
center) field component is enhanced, for polar and azimuthal
components of local field we have $G_{\rm max}<1$. The interval
$\kappa_{\rm max}(\o_{\rm av})=3\dots 10$ corresponds to the
characteristic interval of distances between the molecule and the
nanoparticle surface (usually $2-3$ nm)
\cite{SERS_exp_add,SERS_exp_add1}. In random arrays of
nanoparticles their electromagnetic interaction changes the
electromagnetic SERS gain $G_{\rm max}$ not significantly.

Some other mechanisms (chemical adsorption of the metal interface
that modifies the molecule polarizability \cite{SERS_exp_add} and
some quantum effects \cite{SERS_exp_add1}) bring additional
increase of the molecule dipole moment at Raman frequencies. As a
result, SERS for a molecule located at same distances from the
surface of a spherical nanoparticle corresponds to the total
amplitude gain $G^{\rm tot}_{\rm max}\sim (1-3)\cdot 10^3$ or
total intensity gain $GI^{\rm tot}_{\rm max}\sim 10^{(6-7)}$
\cite{SERS_exp2,SERS_exp_add}. Here the superscript means that in
coefficients $G^{\rm tot}_{\rm max}(\o_{\rm av})$ and $GI^{\rm
tot}_{\rm max}(\o_{\rm av})$ not only electromagnetic mechanisms
of SERS are taken into account.

Specially grown nanowires or nanodisks instead of spheres or
ellipsoids particles improve the result for the maximal
electromagnetic gain $10^1\dots 10^2$ times for points near the
wires ends or disk edges (e.g. \cite{Goudon}). Tremendous
electromagnetic gain $GI_{\rm max}\sim 10^{12}$ can be obtained
for molecules located within small ($0.5-2$ nm sized) gaps between
paired nanoparticles (nanospheres, nanorods, bow-tie arms) as well
as in similar gaps between almost touching particles of plasmonic
nanoclusters \cite{SERS_exp_huge,SERS_molecules}. That allowed one
to apply SERS for detecting separate molecules
\cite{SERS_molecules}.

A drawback of these exciting variants of SERS is the extreme
locality of the huge field enhancement
\cite{SERS_huge1,SERS_huge2}.
Thus, this method practically requires to direct a molecule under
detecting to a selected point. In \cite{SERS_exp_interaction} the
practical importance of the highest possible \emph{averaged}
electromagnetic SERS gain $GI_{\rm av}$ was stressed. The
averaging should be done over the effective domain occupied by
metal nanoparticles or nanocorrugations (minus the volume occupied
by the metal) \cite{SERS_exp_interaction}. In
\cite{SERS_exp_interaction} values $GI_{\rm av}\sim 10^7$ were
theoretically engineered in a regular array of exactly touching
cylindrical nanocorrugations on the silver half-space. Since
\cite{SERS_exp_interaction} the progress in the design of high
$GI_{\rm av}$ has been significant. The mechanism of the high
$GI_{\rm av}$ is related to the electromagnetic interaction in
regular arrays of plasmonic nanoelements. In regular arrays values
of $GI_{\rm av}$ can attain $GI_{\rm av}=2\cdot 10^{11}$
\cite{Genov} that is accompanied by the maximal gain $GI_{\rm
max}\sim 10^{14-15}$ \cite{SERS_huge_array1,SERS_huge_array2} at
the crevices between the corrugations
. However, these amazing values of the averaged gain require the
extreme precision of nanofabrication. Statistical deviations in
the array geometry of the order of one Angstrom lead to its
dramatic decrease from $GI_{\rm av}\sim 10^{11}$ to $GI_{\rm
av}\sim 10^{(6-7)}$ \cite{Genov}.

The idea of the present paper is to show the way to very high
values of $GI_{\rm av}$ using random arrays of metal nanoparticles
on the dielectric core to be fabricated by the self-assembly. Very
high $GI_{\rm av}$ theoretically results from the combination of
SERS with cavity-enhanced Raman scattering. The last one is an
important direction of the modern literature. The strong field
enhancement holds around the points corresponding to the
whispering gallery (WG) modes maxima inside the optical
microcavity \cite{1,2,3,4,5,6,7,8} that also results in the high
Raman gain (defined for these structures in a different way than
in SERS). Definitely, this method is not applicable for detecting
the separate molecules since the field of WG modes is concentrated
inside the solid cavity. The cavity-enhanced Raman scattering
effect is practically thought as promising for microlasers and is
considered usually as the stimulated Raman scattering (SRS) effect
\cite{Raman_microlaser}. In SRS the incident light is used for
pumping the WG states of the microcavity which are excited at
Raman frequencies of the cavity material.

Arrays of touching microcavities were suggested and studied in
work \cite{CERS2007} with the purpose to enhance the local field
around the contact points. This approach combines some features of
SERS (sensing of molecules in the host medium) and cavity-enhanced
Raman scattering (WG modes at $\o_R$). However the result for
$GI_{\rm av}$ (defined in this case the same way as in SERS) is as
modest as $GI_{\rm av}=7$ \cite{CERS2007}. Notice that the
electromagnetic coupling of an optical microcavity (their typical
radiation quality is as high as $Q\sim 10^{(8-10)}$) to outer wave
fields is negligibly weak. Thus, the simple pumping by incident
waves is not efficient \cite{3,4,5,6} and in cavity-enhanced Raman
scattering schemes one uses special coupling elements (prisms or
waveguides) with wave leakage between them and cavities
\cite{3,4,5,6,7,8,CERS2007}.

\begin{figure}[btp]
\begin{center}
\includegraphics[width=15cm]{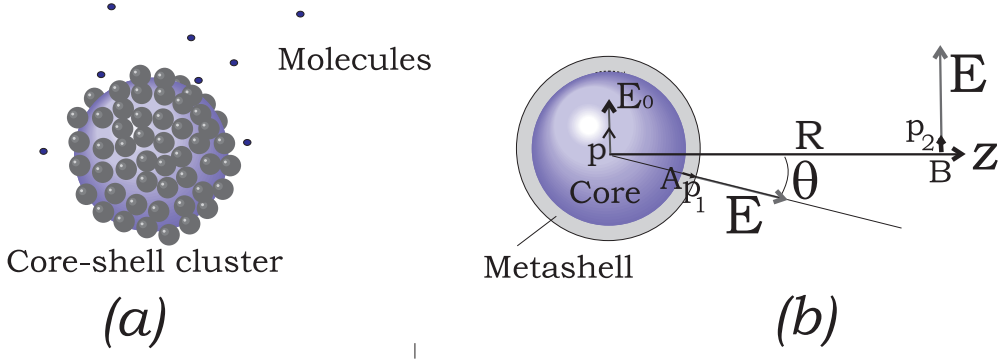}
\caption{(Color online) (a) -- A silica micron- or submicron-sized
sphere (or a cylinder) covered with metal nanoparticles. (b) --
The same structure presented as a core-shell cavity. A point
dipole $p_1$ located in the metashell (the radial component of the
molecule dipole moment) creates at the observation point the
 which equals to the
radial component of the field produced at the center of the dipole
$p_1$ by an auxiliary dipole $p_2=p_1$ placed at the observation
point. $E_0$ is the amplitude of the field of the same auxiliary
dipole in the absence of the core-shell cavity.} \label{figa}
\end{center}
\end{figure}

The goal of the present paper is to show that using a layer of
silver nanoparticles (e.g. spheres) randomly distributed on the
silica core of submicron or micron size as shown in Fig.
\ref{figa} (a) one can significantly improve $GI_{\rm av}$
compared to the conventional SERS based on the same nanoarray on
the planar substrate. Moreover, using such core-shell clusters one
increases the effective volume where the field is enhanced. Such
"templated nanoparticles" (in the terminology of works
\cite{Ku1,Ku2}) can be prepared with existing technologies
\cite{Ku1,Ku2,46} which allow one to obtain also the cores with
high precision of the diameter \cite{Liang}. The technology of
preparing dense one-layer and even multilayer arrays of these
"templated nanoparticles" (or "templated microparticles") on a
dielectric substrate is described in \cite{Ku1,Ku2}. Compared to a
a conventional SERS scheme with plasmonic nanoarray covering a
planar substrate the use of spherical "templated particles" gives
the gain in the total effective volume of the nanoarray
approximately equal to $2\pi$. The use of cylindrical cores gives
the volume gain close to $\pi$. This fact was noticed in works
\cite{Ku1,Ku2} where the SERS in such core-shell nanoclusters was
experimentally studied. The experimentally demonstrated averaged
(over the metashell volume) Raman gain was nearly the same as in
conventional SERS with the same nanoarray on the planar substrate.
The best known result $GI^{tot}_{\rm av}\sim 10^5$ has been
obtained in \cite{46}. Below we predict much better results for
optimized parameters of "templated particles".

In the present paper we consider "templated particles" with
spherical or cylindrical geometry. The metashell couples the
cavity with the host medium and the WG modes can be efficiently
excited by an incident plane wave. Our design goal is to engineer
the cavity resonances so that the WG modes would be concentrated
inside the metashell. The main design parameter is the core radius
$a$. Though the metashell is the same plasmonic nanoarrays which
is used in conventional SERS (on planar substrates), the averaged
gain significantly improves due to the WG resonance.

The metashell can be presented with high accuracy as a layer of an
effective metamaterial shown in Fig. \ref{figa} (b)
\cite{Pastoriza,Rock,PRB}. In works \cite{Rock} the homogenization
model of the spherical metamaterial samples with radius $a=20-50$
nm (diameters of nanoparticles $d=3-6$ nm) was validated by
additional calculations. In \cite{PRB} the same was done for a
metashell on a silica core. In \cite{Pastoriza} the high accuracy
of the homogenization model for the metashell was confirmed by
measured optical spectra where diameters of nanoparticles was
either $d=15$ or $d=20$ nm and the core radius was $a=135$ nm.
These results allow us to use analytical calculations based on the
known solutions of two plane-wave diffraction problems: a
concentric layered sphere \cite{AK1} and a concentric layered
cylinder \cite{AK2}. Explicit expressions for the field in the
inner (core), intermediate (shell) and outer (host medium) regions
can be found in \cite{AK11} for a sphere and in \cite{AK22} for a
cylinder (these papers were used for testing the Matlab codes).
Both Maxwell Garnett (as in \cite{PRB}) and Bruggeman (as in
\cite{Pastoriza}) models were used for the metashell of silver
nanospheres. The chosen design parameters of the metashell (the
nanosphere diameter $d=6$ nm and the averaged gap between
nanospheres $\delta=1$ nm) correspond to the intervals where the
Maxwell Garnett and Bruggeman models give the same values of the
metashell effective permittivity $\va$. The host medium
permittivity was assumed to be free space. The permittivity of
silicon was taken from \cite{Taff} and that of silver from
\cite{Johnson}.

Calculations of the electric field showed that the WG modes
excited by the incident plane wave concentrate inside the
metashell at the blue edge of the collective plasmon resonance of
the metashell (where its permittivity is close to $0$).

First, let us prove that the formula $G_{\rm av}=\kappa_{\rm
av}(\o_e) \chi_{\rm av})(\o_R)\approx \kappa_{\rm max}^2(\o_{\rm
av})$ or $GI_{\rm av}(\o_{\rm av})=\kappa_{\rm av}(\o_{\rm av})^4$
still holds for "templated particles" when $\o_e\approx
\o_R\approx \o_{\rm av}$. Let a molecule be located at a point $A$
inside the metashell of the spherical core. The radial component
of its dipole moment at the Raman frequency is denoted in Fig.
\ref{figa} (b) as $p_1$. Two other components of the dipole moment
are not significant for SERS in spherical "templated particles".
This dipole creates at an arbitrary chosen observation point $B$
located in the far zone the $\theta$-polarized field of complex
amplitude $E(B)$ that we can present in the form $E(B)=\chi(A) p
F(R)$. Here $p F(R)$ ($F$ is the known function) expresses the
field produced at point $B$ in absence of the "templated particle"
by the $y$-oriented dipole $p=p_1$ located at the same distance
$R$ from $B$ as the "templated particle" center (see Fig.
\ref{figa}). From reciprocity $E(B)$ equals to the radial
component of the field ${\bf E'}(A)$ produced at point $A$ by an
auxiliary $y$-oriented dipole $p_2=p_1=p$ located at $B$. Since
the dipole $p_2$ is located very far from the "templated particle"
the incident field from which the radial field $E(A)\equiv
E'_r(A)$ results can be approximated as the field of a plane wave
with amplitude $E_0=F(R)p$ at the sphere center. By definition the
local field amplitude enhancement is equal to $\kappa(A) \equiv
|E(A)/E_0|$, where as we have seen $E(A)$ results from the plane
wave refraction. Therefore the amplitude radiation enhancement due
to the presence of the "templated particle" defined as
$|\chi(A)|\equiv |E(B)/pF(R)|$ (i.e. with respect to the same
molecule located at the same distance $R$ from the observation
point as the "templated particle" center) is equal to
$|\chi(A)|\equiv |E(B)/p F(R)|=|E(A)/E_0|=\kappa(A)$. Since this
result holds for an arbitrary point $A$, the averaged radiation
and local field enhancements are also equivalent. So, the
enhancement of the molecule radiation at the Raman frequency in
presence of the spherical "templated particle" is equal to the
local field enhancement (with respect to an incident plane wave)
in presence of the same nanoparticle in the same place. The same
speculation can be done for other field components, for a
cylindrical "templated particle", and for any other finite
structure modifying the local field.

To find $G_{\rm av}$ we have to average the intensity $|E(A)|^2$
over the metashell volume (for spherical and cylindrical cores it
is a simple numerical integration of analytical expressions). Here
$E(A)=\kappa(\o_{\rm av},A)$ is the complex amplitude of the
radial component of electric field when the "templated particle"
is impinged by a plane wave of unit amplitude. The intensity Raman
gain is equal to $GI_{\rm av}=G_{\rm av}^2$.

\begin{figure}
\begin{center}
\subfigure[][]{\label{Y1}\includegraphics[width=7cm]{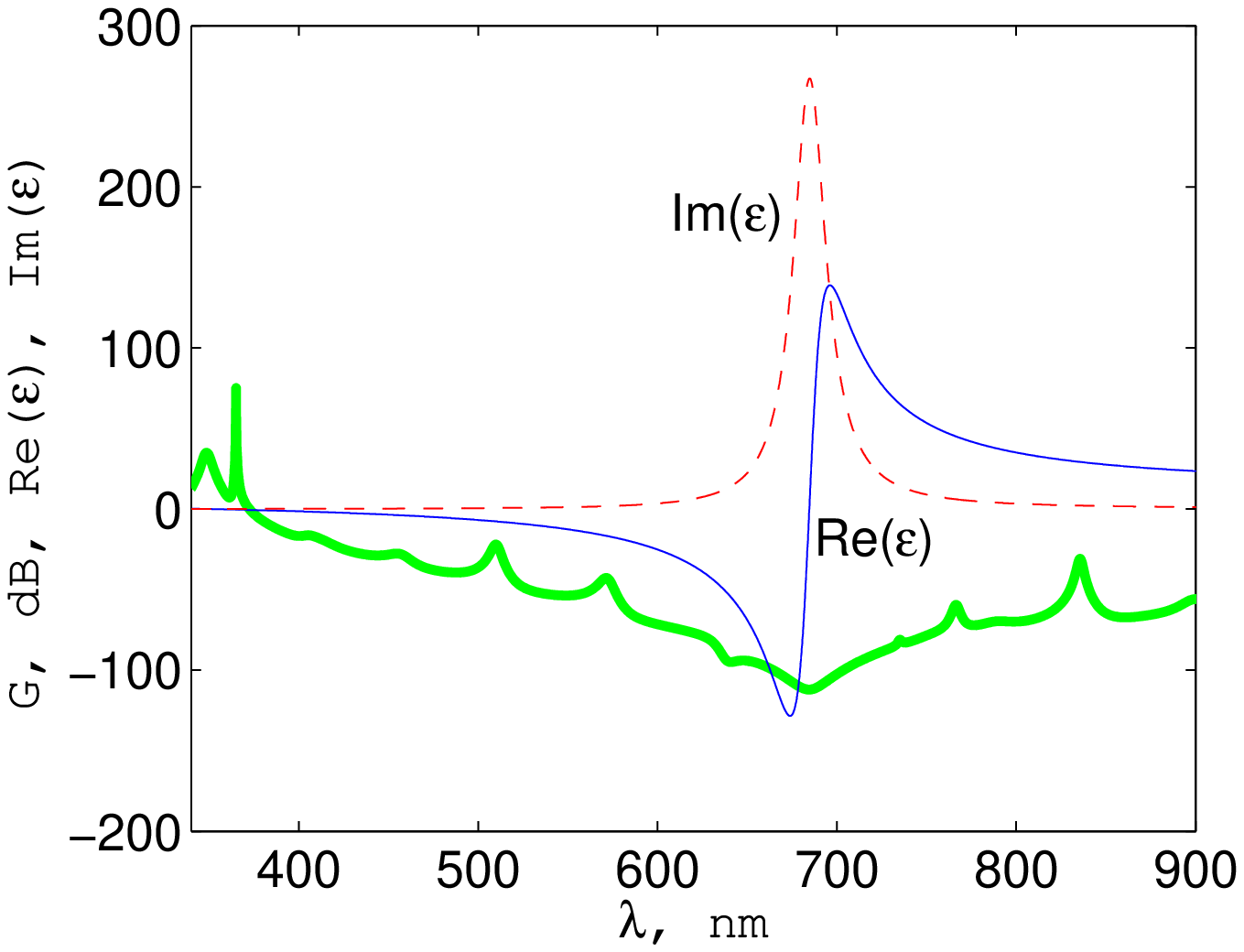}}
\subfigure[][]{\label{Y2}\includegraphics[width=7cm]{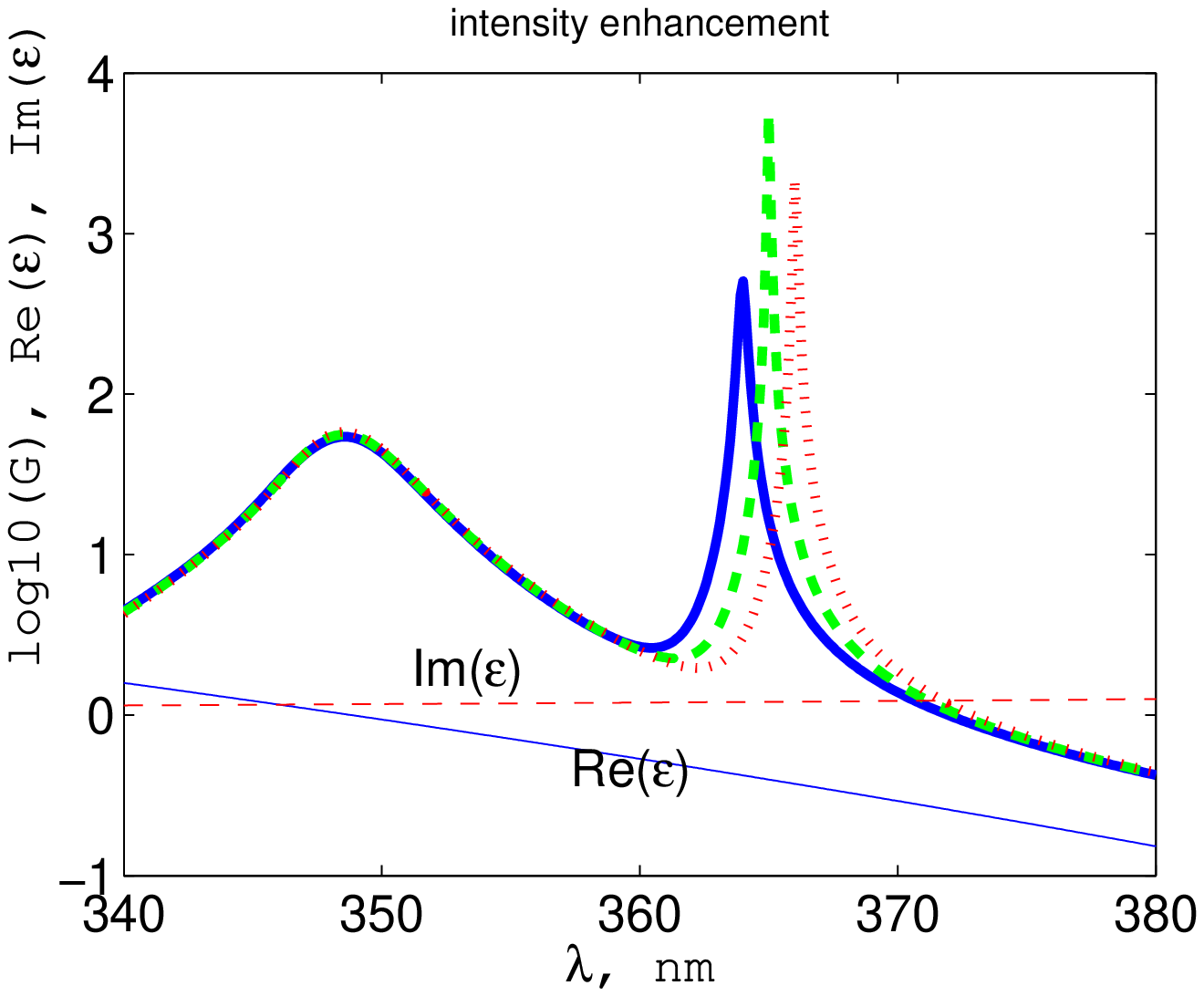}}
\caption{(Color online)
 \subref{Y1} -- Wide-band frequency plot of the averaged local field intensity gain $G$ (in dB) for spherical core with radius $a=323$ nm
 and of the complex permittivity of the metashell.
\subref{Y2} -- Same as \subref{Y1} in a narrow band for three
values of the core radius $a=322$ nm (solid line), $a=323$ nm
(dashed line), $a=324$ nm (dotted line).
 }
\label{figa1}
\end{center}
\end{figure}

The result for $G_{\rm av}$ (in dB) versus light wavelength for
the spherical "templated particle" with the core of radius $a=323$
nm is shown in Fig. \ref{figa1} (a) together with the complex
permittivity of the metashell. At the collective plasmon resonance
of the metashell, the losses are very high and $G_{\rm av}\ll 1$.
The "useful" WG resonance (when the field concentrates inside the
shell) corresponds to the azimuthal and polar numbers of the WG
mode $L=1$ and $N=9$, respectively. It holds at $\lambda=366$ nm
where $\va=-0.1+i0.16$ and $G_{\rm av}= 5784$. It corresponds to
the averaged intensity Raman gain $GI_{\rm av}\sim 4\cdot 10^7$
and to the cavity optical quality  $Q\approx 5\cdot 10^6$. The
presence of the metashell broadens the band of the WG resonator.
Cores with $a=322$ and $a=324$ nm still correspond to $GI_{\rm
av}\sim 3\cdot 10^7$, as we can see from Fig. \ref{figa1} (b),
where (${\rm log}_{10} G_{\rm av}$ is shown in more details for
three values of the Si sphere radius. This result shows the
possible tolerances in the cavity fabrication.

It had been expected that the higher-order WG resonances (when
$a\ge 1\ \mu$m) should have strongly increased the gain since the
optical quality of cavities usually grows along with the resonance
order. However, for "templated spheres" it is not so. The next
"useful" WG resonance corresponds to $a=611$ nm ($L=1,\ N=18$),
however $GI_{\rm av}\sim 3$ does not improve since the WG
resonance shifts to lower frequencies where ${\rm Im}(\va)$
increases and the losses reduce the gain.

\begin{figure}
\begin{center}
\subfigure[][]{\label{Y11}\includegraphics[width=7cm]{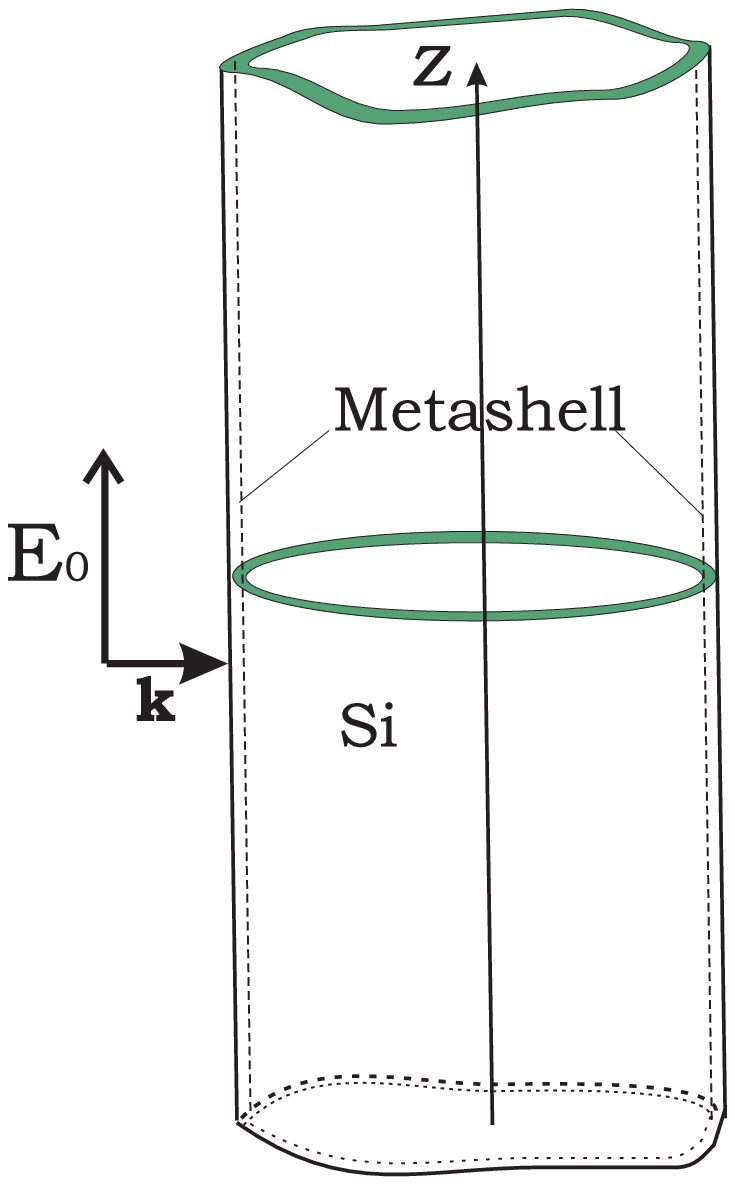}}
\subfigure[][]{\label{Y22}\includegraphics[width=7cm]{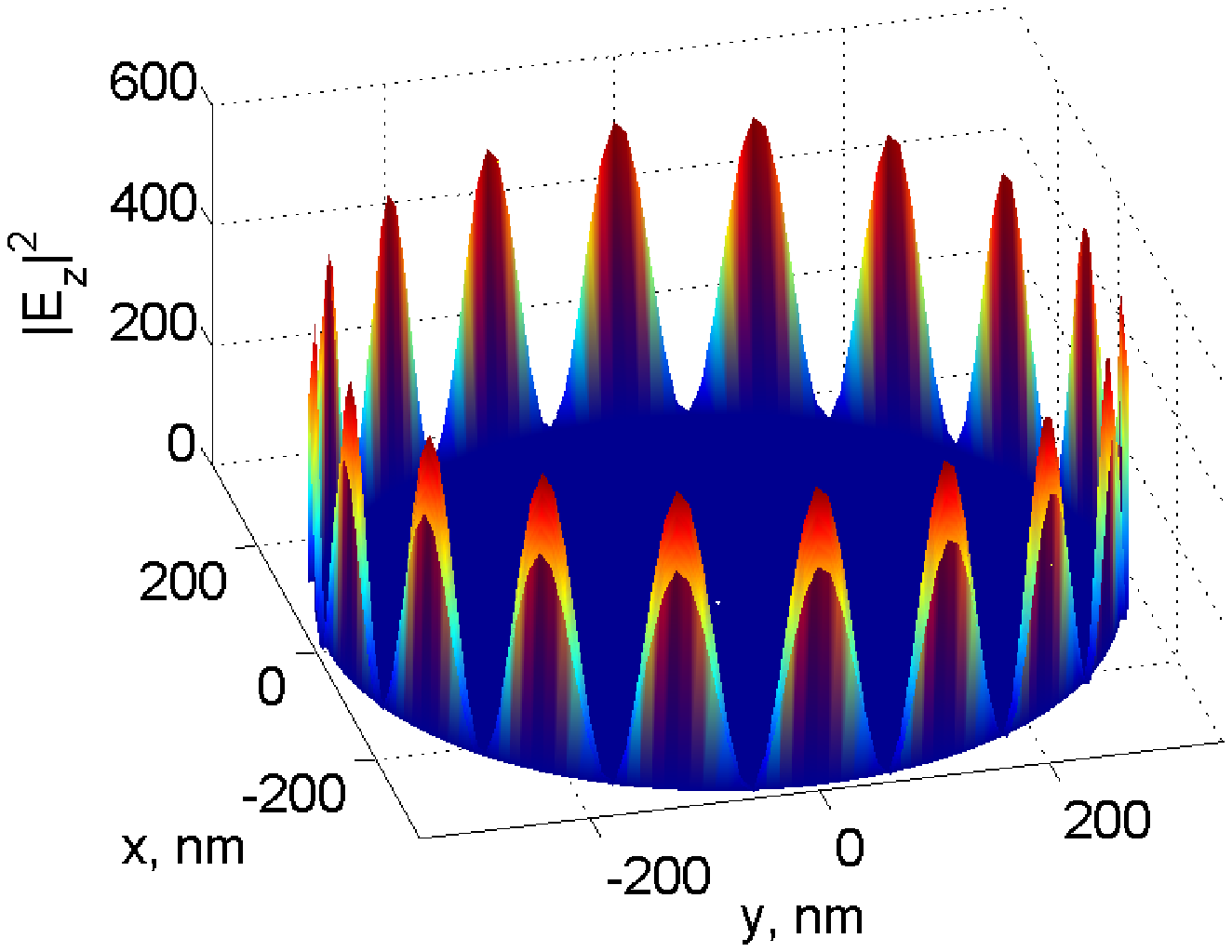}}
\caption{(Color online)
 \subref{Y11} -- Cylindrical silica core with the same metashell impinged by incident plane wave.
\subref{Y22}-- Electric intensity distribution in the structure
with $a=337.5$ nm at $\lambda=353$ nm and $G_{\rm av}=202$.
 }
\label{figa2}
\end{center}
\end{figure}

For cylinders this blue shift is absent and the gain grows versus
the order of the "useful" resonance. The case shown in Fig.
\ref{figa2} (a), i.e. the normal incidence of an axially polarized
plane wave to an infinitely long cylinder of radius $a$ covered
with the same metashell was studied. There is only one component
of the electric field $E=E_z$. In Fig. \ref{figa2} (b) the
distribution of the field intensity over the structure with
$a=337.5$ nm at $\lambda=353.5$ nm (the mode azimuthal number
$L=9$) is shown. The localization of the field inside the
metashell is clearly seen. One can see that at this frequency
$G_{\rm max}=|E_z|^2_{\rm max}$ exceeds $500$. However, the
numerical averaging gives $G_{\rm av}=202$ i.e. $GI_{\rm av}\sim
4\cdot 10^4$.

\begin{figure}
\begin{center}
\subfigure[][]{\label{Y111}\includegraphics[width=7cm]{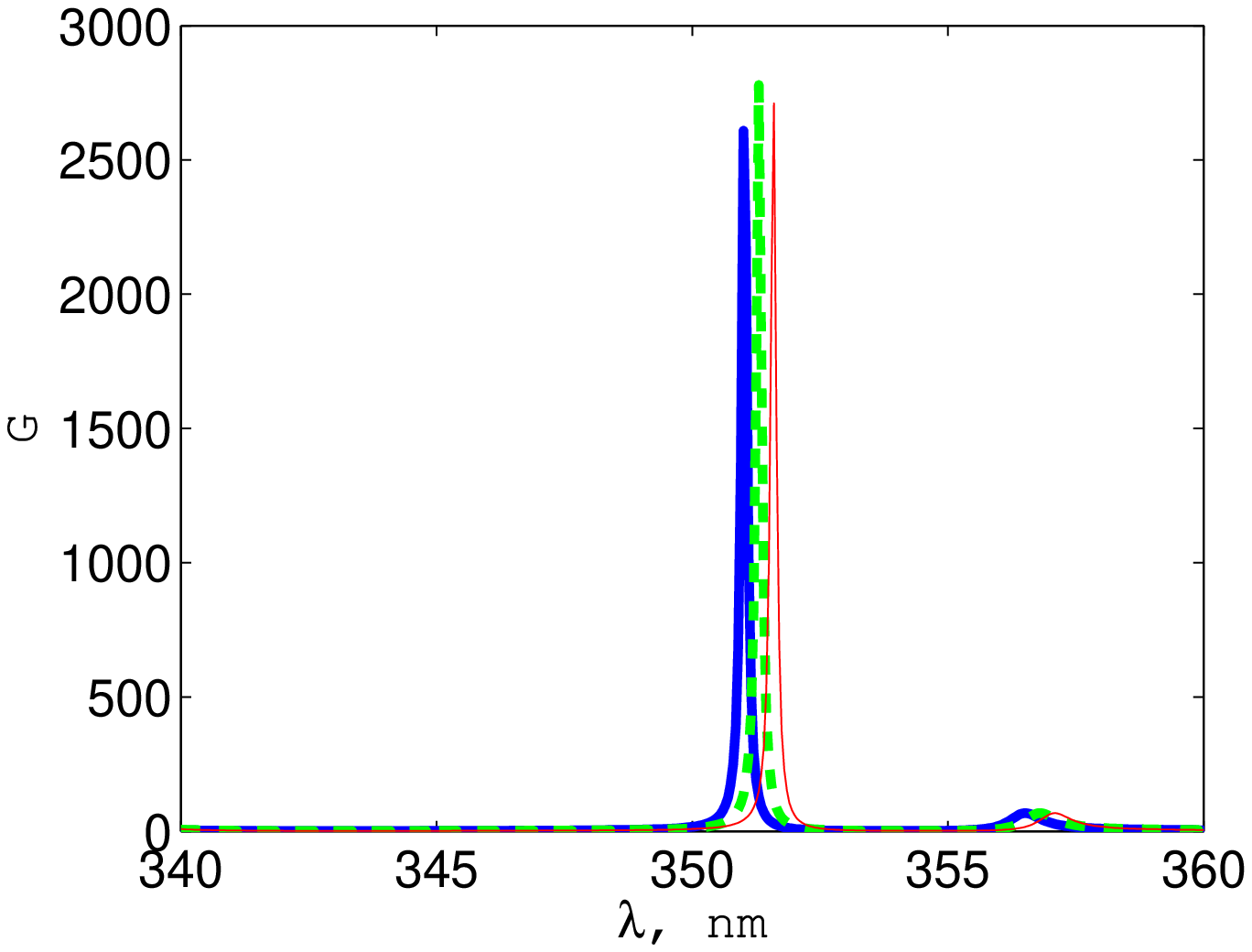}}
\subfigure[][]{\label{Y222}\includegraphics[width=7cm]{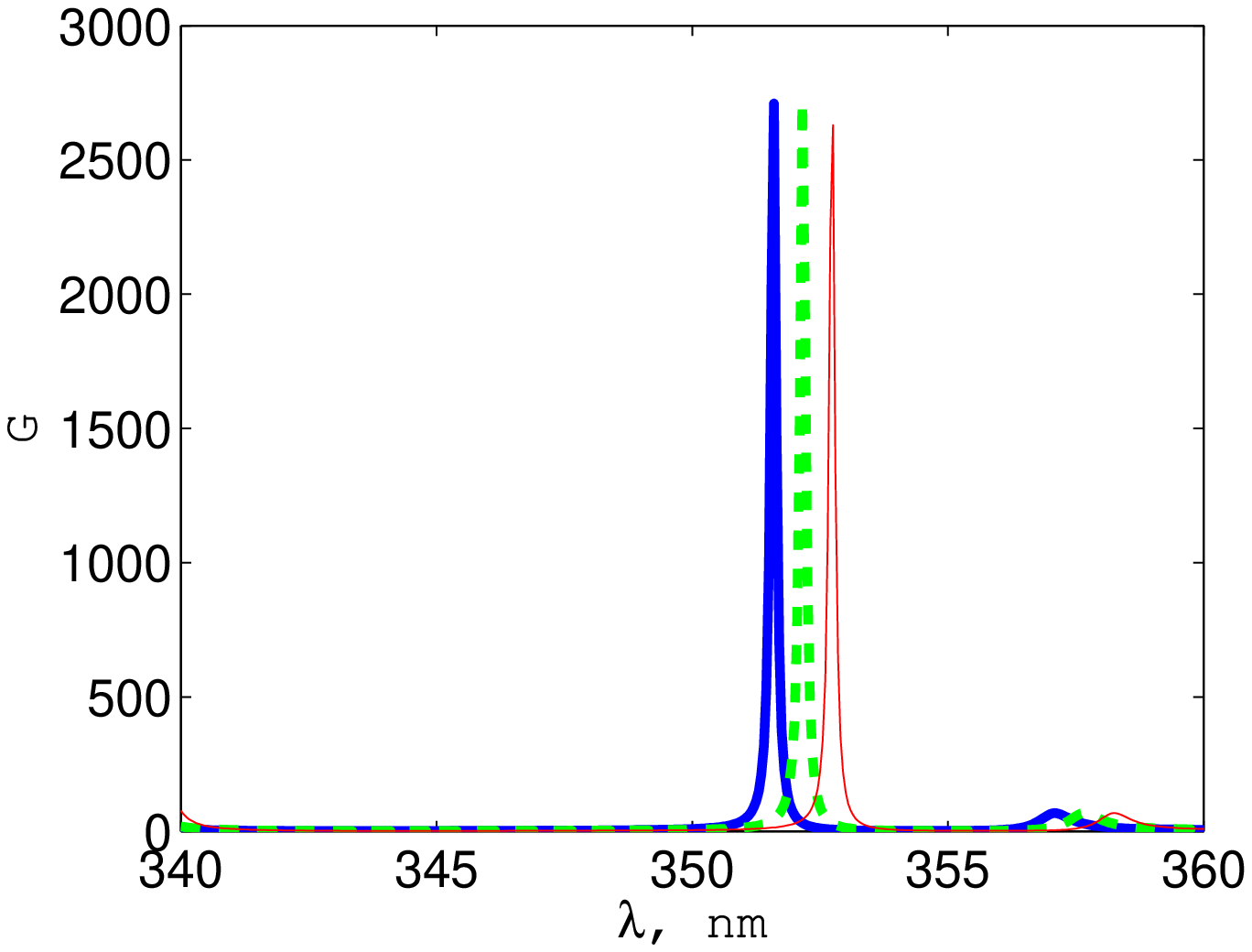}}
\caption{(Color online)
 \subref{Y111} -- Frequency dependence of $G_{\rm av}$ for three values of the cylinder radius $a=615.5,\ 616,\
616.5$ nm (thick solid line, thick dashed line and thin solid
line, respectively)).
\subref{Y222} -- Same as  \subref{Y111} where the metashell is an
abstract dispersionless material with permittivity  $\va_{\rm
const}=-0.08+i0.156$.
 }
\label{figa3}
\end{center}
\end{figure}

Fig. \ref{figa3} (a) shows the frequency dependence of $G_{\rm
av}$ for three values of the Si cylinder radius $a=615.5,\ 616,\
616.5$ nm corresponding to the WG modes also concentrated within
the shell (the resonance order $L=18$). Then the result is much
better $G_{\rm av}=2500\dots 2700$ ($GI_{\rm av}\sim 10^7$) within
the band $\lambda=\lambda_{\rm av}\pm 1$ nm, where $\lambda_{\rm
av}=352$ nm. In Fig. \ref{figa3} (b) the results are shown for an
abstract metashell with dispersionless permittivity $\va_{\rm
const}=-0.08+i0.156$ which is equal to $\va (\lambda_{\rm av})$.
It is clear that the dispersion of the metashell permittivity has
no impact but for the highest gain one has to adjust $\va$ (in the
region of $|\va|\ll 1$). The same structure at same frequency
gives $G_{\rm av}=3\cdot 10^4$ for $\va_{\rm const}=-0.01+i0.01$.

Further increase of the cylinder radius allows us to obtain the
higher gain. The result corresponding to the WG resonance of the
order $L=32$ is $G_{\rm av}(\lambda=353{\rm nm})\approx 10^5$. In
this case the band of the WG resonance becomes as narrow as
$10^{-3}$ nm and the optical quality is of the order $Q\sim 10^7$.
Notice, that very high values of $GI_{\rm av}$ correspond to very
narrow resonance band. Then the excitation frequency and the Raman
frequency can belong to this resonance band only if the Raman
shift is very small. Otherwise the approximation $\o_e\approx
\o_R\approx \o_{\rm av}$ is not anymore valid, and the result
$|\chi(\o_{R})|=\kappa(\o_e)$ does not hold. The "template
particles" with $a\ge 1\ \mu$m have at the "useful" WG resonance
the optical quality of the order $Q\ge 10^{9}$ and huge Raman gain
$GI_{\rm av}(\lambda_{\rm av})\sim 10^{12}$. However this gain can
be applied only for molecules with very small Raman shift (less 1
cm$^{-1}$ for wavenumbers). Therefore we do not consider so big
"templated particles" in more details.

To conclude: it is theoretically demonstrated that the WG
resonances in core-shell "templated particles" whose effective
shell permittivity is close to zero are very promising for SERS.
These structures combine useful features of SERS and
cavity-enhanced Raman scattering. Exact analytical calculations
were also done (for both spheres and cylinders) covered with a
solid silver shell of same thickness $d=6$ nm. The WG resonances
at which the field of WG modes is strongly localized (concentrated
at the inner interface between the core and the shell) were found.
These resonances demonstrated tremendous $G_{\rm av}$ that attains
even for submicron $a$ values $G_{\rm av}\sim 10^{6}$ ($GI_{\rm
av}\sim 10^{12}$). This effect can be applied in prospective SRS
schemes.

\end{document}